# 100 years of plastic: using the past to guide the future

**Authors:** Chao Liu[1]*†, Roland Geyer[2]*†, Shanying Hu[1]†

**Affiliations:**
[1]Department of Chemical Engineering, Tsinghua University; Beijing, 10084, China.
[2]Bren School of Environmental Science and Management, University of California Santa Barbara; Santa Barbara, CA, 93106, USA.

*Authors contributed equally
†Corresponding authors: cliu20@mails.tsinghua.edu.cn, geyer@bren.ucsb.edu, hxr-dce@mail.tsinghua.edu.cn

**Abstract:**
Robust and credible material flow data are required to support the ongoing efforts to reconcile the economic and social benefits of plastics with their human and environmental health impacts. This study presents a global, but regionalized, life cycle material flow analysis (MFA) of all plastic polymers and applications for the period 1950-2020. It also illustrates how this dataset can be used to generate possible scenarios for the next 30 years. The historical account documents how the relentless growth of plastic production and use has consistently outpaced waste management systems worldwide and currently generates on the order of 60 Mt of mismanaged plastic waste annually. The scenarios show that robust interventions are needed to avoid annual plastic waste mismanagement from doubling by 2050.

**Main Text:**
The year 1950 has been proposed as both the beginning of the Anthropocene, a new geological era in which humanity dominates geological processes, and the onset of a sharp increase in human population and economic activity, known as the Great Acceleration (*1*). The same year also marks the start of mass production of synthetic polymers, or plastics (*2*). The low economic cost and technical versatility of plastics have resulted in large and sustained growth in their global production and use (*2,3*). Global mass accounts suggest that cumulative plastic production had reached twice the combined mass of all living terrestrial and marine animals in 2015, and that all human-made mass had exceeded all living biomass in 2020 (*4*).

Lack of degradation, bio-assimilation, and suitable waste management infrastructure has led to plastics not just being everywhere in our lives, but also everywhere in the environment. There is increasing concern about the human and environmental health consequences of relentless growth in macro and micro plastic pollution (*5-10*). After decades of "anonymity in ubiquity", plastics have finally come under intense scrutiny by scientists, policy makers, and the public at large (*11*). Across the world, efforts are underway to reconcile the economic and social benefits of plastics with their human and environmental health impacts. One prominent occasion is the legally binding international instrument known as the Global Plastics Treaty, which is currently being negotiated through the Environmental Nations Environment Programme.

A growing number of plastic flow accounts has become available in support of these management and policy efforts. Most global studies have focused on the generation and fate of plastic waste (*12-15*). Studies that investigate and model the entire life cycle of plastics, on the other hand, are typically done on the national level and often limited to specific polymers and just one year (*16-35*). The research presented here adds to the existing literature through developing a global, but regionalized, account of the entire life cycle of all plastic polymers and applications for the period 1950-2020. This approach is a necessary step towards being able to compare and reconcile global and national accounts of plastic production, use, and waste management. It can also be used to investigate equity issues within the global plastic system. The study concludes by illustrating how the introduced comprehensive historical dataset can be used to generate plausible future plastic scenarios and thus inform global and regional plastic policy development.

This study uses material flow analysis (MFA), which is based on the mass balance principle, and follows annual production flows of plastic resin, fibers, and additives downstream through their life cycle from conversion and use, to waste generation and end-of-life (EOL) management (*2*). The MFA is global, and the world is divided into four regions: China, North America (Canada, USA, and Mexico), EU30 (European Union plus UK, Switzerland, and Norway), and the rest of the world (RoW). The historical part covers the period 1950 to 2020. The regionalized nature of the MFA requires to consider trade at all stages of the plastic life cycle, including waste management. It also necessitates region-specific data and models for all life cycle stages. This results in four trade-linked regional plastic MFAs, which cover the first 70 years of global plastic mass production and consumption. All polymer types and consuming sectors are accounted for. Plastic flows are characterized by 9 polymer types and 8 consuming sectors.

The comprehensive historical plastic flow data from 1950 to 2020 is used to generate several possible scenarios for the next 30 years of plastics until 2050. Key inputs to generate these projections are historical and forecast population and GDP data for the four world regions and the 100 year timeframe (1950-2050). The business as usual (BAU) scenario assumes that the trends observed during the last 10 years continue for the next 30 years. Two additional scenarios are created by modeling plausible deviations from the BAU trends. Rather than making predictions, the objective of the scenario analysis is to illustrate how the comprehensive global historical plastic flow dataset can be used to inform and guide plastic management decisions and policy analysis at this crucial point in time for humanity's relationship with plastics.

**The past 70 years**

Annual global virgin plastic production went from 2 Mt in 1950 to 473 Mt in 2020. With a few exceptions, such as the oil crises and the Great Recession, it has been following a remarkably stable growth trend: a quadratic fit generates an R squared of 99.8%. Cumulative virgin plastic production now exceeds 10 billion metric tons, half of which was produced just in the last 14 years. At regional scale, 39% of 2020 annual production took place in China, 16% in North America, and 13% in the EU30. 30 years ago, the production shares were 3%, 26% and 33%, respectively. For RoW, this share has been fluctuating between 29% and 44% in the last 50 years. Shown as relatively flat curves in the past 15 years (Fig. 1A), European and North American production ranged from 63 to 78 Mt per year. Since the start of the new millennium, Chinese annual production increased about tenfold. Resins, synthetic fibers and additives account for 77%, 16% and 7% of current plastic production. The share of synthetic fibers continues to grow. Six polymer types (PE, PP, PS, PVC, PET, and PUR) account for 83% of global resin production.

Growth in global virgin plastic production, and thus consumption, has consistently exceeded global population growth. Between 1950 and 2020, global per capita consumption has been rising steadily from 1 to 60 kg per capita, with a fairly linear trend ($R^2 = 0.98$) (Fig. 1B). At the regional level, after the Great Regression, European per capita consumption has been fluctuating between 139 and 156 kg, while that of North America has been on an upward trend again and stood at 178 kg per capita in 2020. Chinese per capita consumption exceeded the global average in 2008, and reached 124 kg in 2020, while it was only 4kg in 1990. The oil crises and the Great Recession had large impacts on the growth of virgin plastic consumption in North America and EU30, but much less so elsewhere. 137 Mt, or 29% of global virgin plastic consumption in 2020, was for packaging, making this by far the largest plastic application. In joint second place are textiles with 16%, or 76Mt, and building and construction with 17%, or 80 Mt. The next largest sector is household/leisure/sports with 57 Mt, or 12%. With a few exceptions, consumption by sector is fairly similar across the four regions (Fig. 2A). The one regional difference that stands out the most is the large share of textiles in Chinese consumption. Another significant regional difference is the large share of the household/leisure/sports category in North American consumption. Cumulative regional consumption of virgin plastic between 1950 and 2020 is as follows: China 1.9 Gt, North America 2.6 Gt, EU30 2.7 Gt, and 3.5 Gt for the rest of the world.

Between 2005 and 2020, trade between the four world regions more than doubled, from 34 to 74 Mt. Trade in plastics takes place all along the value chain and is divided into three categories here: resins and synthetic fibers, plastic articles, and final products. Trade in plastic resins has the largest volumes and doubled from 16 to 33 Mt. Resins' share in total trade, however, decreased slightly from 49% to 45%. Synthetic fibers experienced 7 Mt of trade in 2005 and reached 13 Mt in 2020. Trade in plastic articles was 7 Mt of trade in 2005, and had increased to 20 Mt by 2020. Trade in plastics contained in final goods has the smallest volume, but experienced the largest growth, from 2 to 8 Mt. To better isolate longer term trends in plastic trade from its significant annual volatility, two 5-year averages are shown here (Fig. 3). On the regional level, interesting patterns emerge. China, for example, had a small trade deficit during 2005-09; net importing 16 Mt of resins and net exporting 15 Mt of plastics in articles, textiles, and goods, per year on average. By 2016-20, this changed to a substantial trade surplus, with 24 Mt of average annual net resin imports and 35 Mt of average annual net exports of plastics in articles, textiles, and goods. North America, on the other hand, was and remains a net exporter of resins and a net importer of plastic in articles, textiles, and goods; with a persistent overall plastic trade deficit.

Global annual generation of plastic waste increased from 5 Mt in 1960 to 412 Mt in 2020. By 2020, global cumulative plastic waste had reached 8 Gt, more than half of which came from only the last 13 years. With 18%, China accounted for the lowest fraction of global cumulative plastic waste. North America's and Europe's contributions were 24% and 27%, respectively. In the year 2020, China and RoW generated the highest annual amounts of plastic waste, which were 126 Mt and 129 Mt. With 74 Mt and 83 Mt, respectively, North America and Europe had lower values. However, the situation was very different at the beginning of the new millennium, when China produced 13 Mt of annual plastic waste, while the other three regions all exceeded 45 Mt. With respect to 2020 waste generation per capita, North America and Europe stood out with 150 kg and 157 kg respectively. In 2020, China's per capita waste generation reached 89 kg, which was still lower than NA's and Europe's values for the year 2000 (111 kg and 95 kg). Throughout the years, RoW has had the lowest waste generation per capita, which has not yet exceeded 25 kg. Primary plastic waste is dominated by packaging (38%), followed by textiles (17%) and household/leisure/sports (14%) (Fig 2A). In all regions, the waste shares of packaging and textiles are higher than their consumption shares.

In this MFA, plastic waste is called mismanaged if it does not enter the formal waste management system to be recycled, incinerated, or landfilled (Fig 2B). All subsequent results are expressed in percent of generated plastic waste. Over the last 70 years, the world has approximately recycled 13% and incinerated 16% of its cumulative plastic waste. The remaining 71% were either landfilled or mismanaged. Formal recycling and incineration are very recent phenomena across the globe. Just the last 15 years account for 84% of all plastic waste that has ever been recycled and 75% of plastic waste that was formally incinerated. Over the last 20 years, China's landfill and recycling rates have varied between 27% and 41%. Formal incineration grew from close to zero to 37%, while plastic waste mismanagement has decreased to an estimated 2% in 2020. Over the same 20 year period, formal plastic waste management of NA was quite stable, with a constant landfill rate of around 72% and only a 3% increase of recycling to 9%. Plastic waste mismanagement decreased from 7 to 2% in 2020. The EU 30 shows the clearest trends of the four world regions. Between 2000 and 2020 the recycling rate went from 9% to 26% and the incineration rate from 26% to 47%. As a consequence, landfill decreased from 57% to 23%, and plastic waste mismanagement was down to 3% in 2020. Accounting for all other world regions makes RoW quite heterogeneous and the search for trends more questionable. Nevertheless, the data show a significant growth of the RoW recycling rate over the last 20 years, from 6% to 17%, but also a persistently high rate of plastic waste mismanagement, which declined from 54% to 44% during the same period.

**The next 30 years**

Three scenarios for the next 30 years were developed to illustrate how the historical plastic flow accounts can be used for policy analysis: S1 is a business-as-usual scenario (BAU) that extends historical trends. S2 assumes that a 60% recycling rate is reached across all sectors in 2050. S3 assumes that a 100% recycling rate is reached in the packaging sector, while all other sector follow the BAU trend. The increased recycling scenarios assume that 75% of recycled plastic displaces virgin production, while the other 25% increase total production and use (*6*).

The next 30 years are set to see dramatic changes under business-as-usual (BAU) trends. Annual plastic production is projected to reach 1.12 Gt in 2050. 75% of this is virgin plastic and 25% is recycled plastic (Fig. 4A). Around the year 2035, annual virgin plastic production starts to deviate from and stay below the long-term quadratic trend mentioned earlier, but is still remains growing. As a result, cumulative virgin plastic production will triple from 10.6 Gt between 1950 and 2020 to 31.3 Gt between 1950 and 2050. Annual per capita consumption of virgin plastic is also projected to increase in all four world regions between 2020 and 2050: from 124 to 243 kg in China, from 178 to 240 kg in North America, from 140 to 185 kg in Europe, and from 25 to 41 kg in the rest of the world. Due to this continuing growth in plastic production and use, cumulative virgin waste generation is set to increase more than threefold from 7.4 Gt in 2020 to 24.4 Gt in 2050. Plastic waste management is also projected to change significantly. The global recycling rate will grow from 21% in 2020 to 30% in 2050. Global incineration rate will tick up from 25% to 28%, while the formal landfill rate will decrease from 39% to 29%. The global plastic waste mismanagement rate will decrease only slightly from 15% to 14%. As a result, the annual amount of mismanaged plastic waste is set to more than double from 63 to 131 Mt. Plastic waste management will remain quite distinct across the four regions. For example, the 2050 recycling rate will reach 38% in 2050 in China and EU30, 25% in RoW, but only 11% in NA. It should be noted here, that BAU projection are predicated on plastic recycling being mostly mechanical, since there has been no significant chemical recycling in the last 70 years. Under BAU, plastic waste mismanagement is projected to further decrease in China, NA, and EU30, but remain stubbornly high at 41% in RoW.

Scenario 2 assumes that all four world regions achieve a 60% recycling rate across all sectors in 2050. It further assumes that this proportionally decreases the other waste management fates. One last assumption is that 75% of recycled plastic displaces virgin plastic, while the remaining 25% increase total plastic consumption (*6*). Alternative assumptions are, or course, possible and should be explored, but this is one possible scenario worth exploring. Due to the circular economy rebound effect of plastic recycling, total annual plastic production is projected to increase to 1.27 Gt in 2050. However, only 660 Mt will be from primary (virgin) production, while the remaining 610 Mt will be from secondary production (recycling) (Fig. 4A). This increase in total production and consumption will also lead to an increase in 2050 annual waste generation; from 0.95 Gt (BAU) to 1.06 Gt. Yet, per scenario definition, 60% of this waste will be recycled, which significantly reduces the other plastic waste fates relative to BAU (Fig. 4B). In particular, it reduces projected annual plastic waste mismanagement in 2050 from 131 Mt (BAU) to 77 Mt. It should be noted that this is still higher than the 2020 value. The only difference between scenarios 2 and 3 is that only the recycling rate of plastic packaging will deviate from BAU and reach 100% in 2050. The outcomes of the two scenarios are very similar, which shows how much more ambitious recycling interventions have to be if only packaging is targeted.

**Using the past to guide the future**

The experience with the Kyoto Protocol on climate change and the subsequent Paris Agreement shows how important detailed and credible historical data are for international treaty negotiations and implementations. The saying that 'you cannot manage what you do not measure' certainly applies here. Our global, but regionalized, life cycle MFA of all polymer types and applications during the first 70 years of plastic mass production is meant as a substantial contribution to the growing need for datasets on plastics. Its completely transparent and open-source nature is designed to help compare and reconcile the increasing number of national plastic MFAs and to support the harmonization of the various emerging accounting methods.

The historical account of the first 70 years of plastic mass production shows that growth in virgin plastic production and consumption has been very stable and persistent. Waste management systems across the world have not been able to keep up with the associated increase in plastic waste generation, and the more recent formal waste management strategies of recycling and incineration have yet to reduce the unsustainable level of mismanaged plastic waste; especially outside of China, North America, and Europe. As a result, and in addition to the emissions associated with plastic production, plastic material itself has now become a major environmental pollutant. The presented plastic flow account also shows that plastic packaging is far from being the only type of plastic application that requires careful management and robust interventions.

The business as usual scenario shows that the projected growth in formal incineration and recycling will not be able to keep up with the projected growth in plastic production and use. However, the two additional exemplary recycling scenarios also show that many other plastic futures are possible and should be explored and hopefully implemented. While robust data and models are a key requirement for solving global environmental problems such as plastic pollution, it is just one of many. What is needed the most is that all stakeholders across all regions of the world come together in a good faith effort to solve the issues at hand. The INC meetings of the global plastics treaty provide a timely and much-needed venue for this.

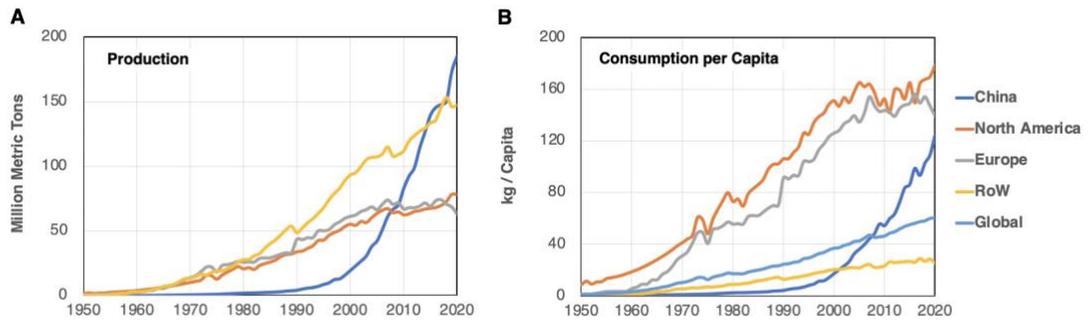

**Fig. 1. Plastic production and per capita consumption between 1950 and 2020.**

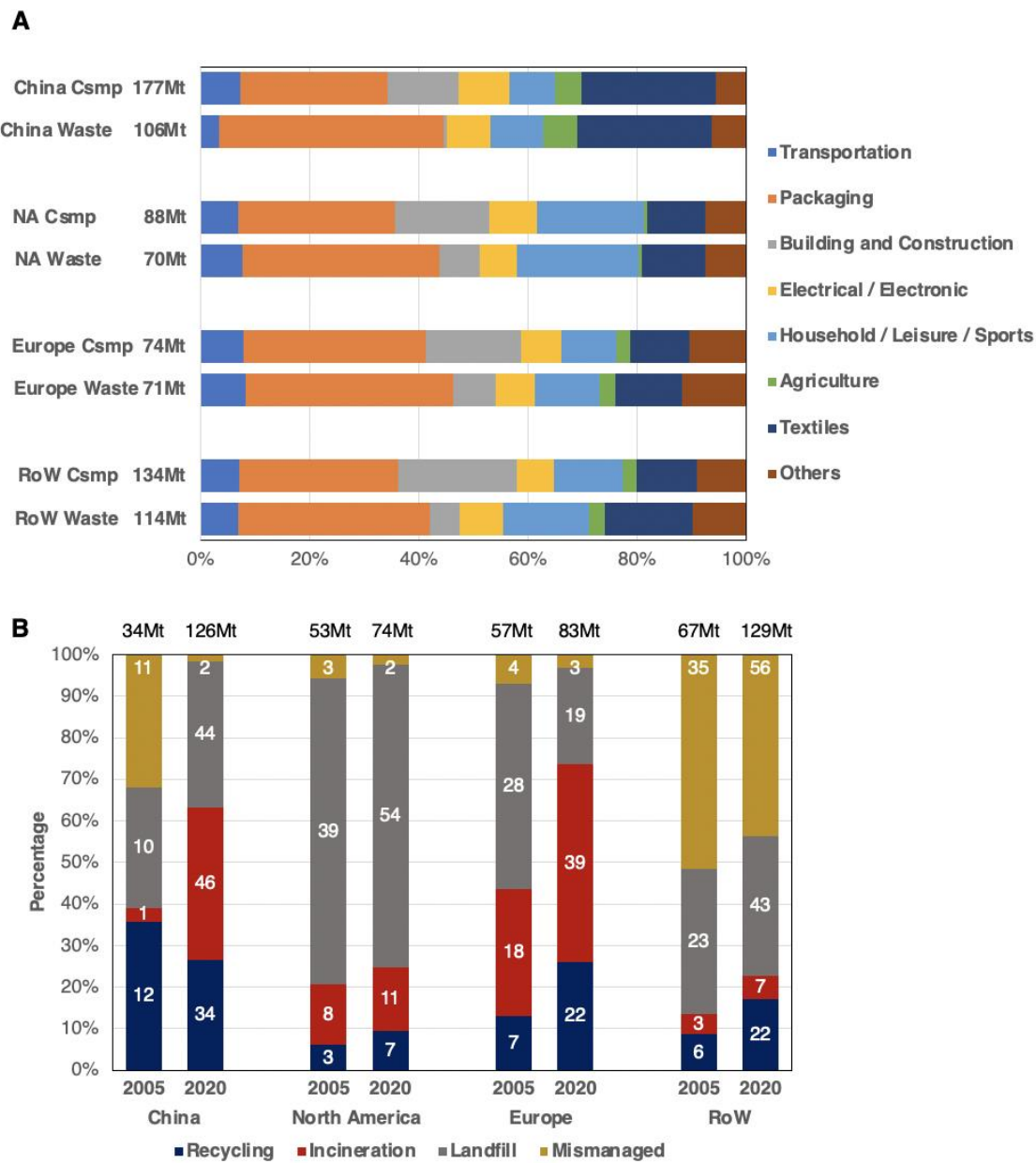

**Fig. 2. Regional plastic consumption and waste distribution by sector and end of life management evolution.** (**A**) 2020 regional plastic consumption and primary waste generation by sector. (**B**) 2020 and 2050 regional plastic end of life treatment (primary and secondary waste), the numbers at the top of the figure indicate the absolute value in Mt.

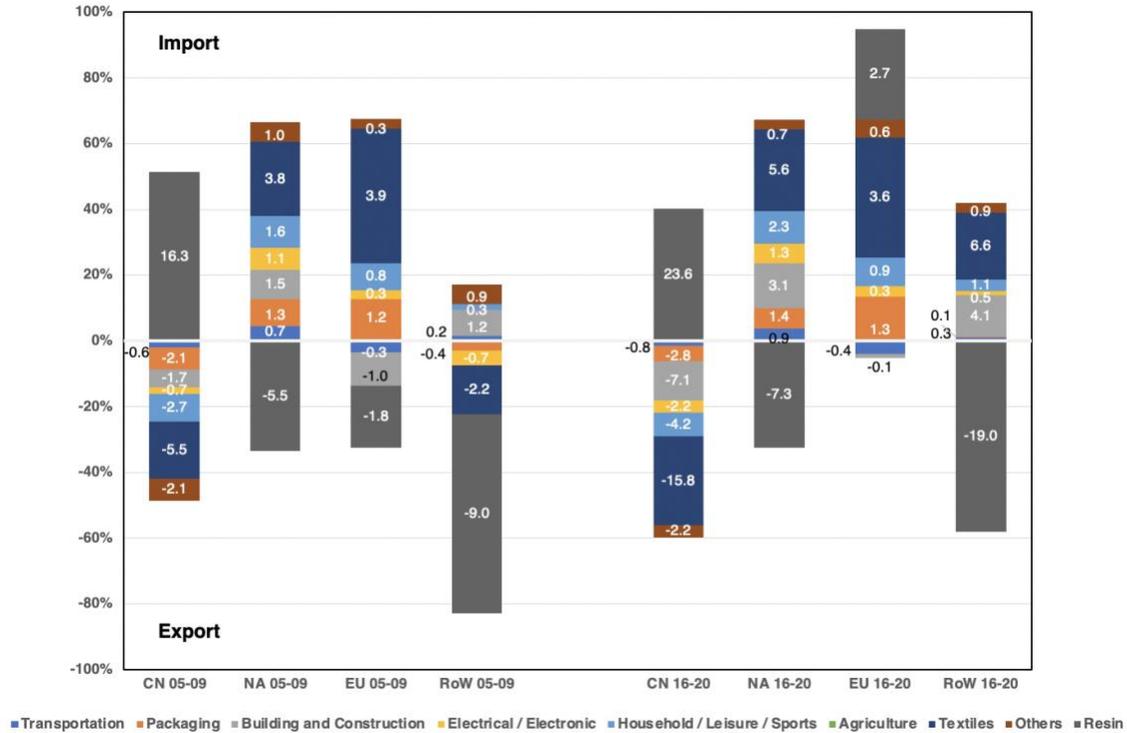

**Fig. 3. Trade by sector.** 2005-2009 average and 2016-2020 average. The height of the bars indicates percent of total trade within the region. Absolute values on the bars are in Mt. CN – China, NA – North America, EU – Europe.

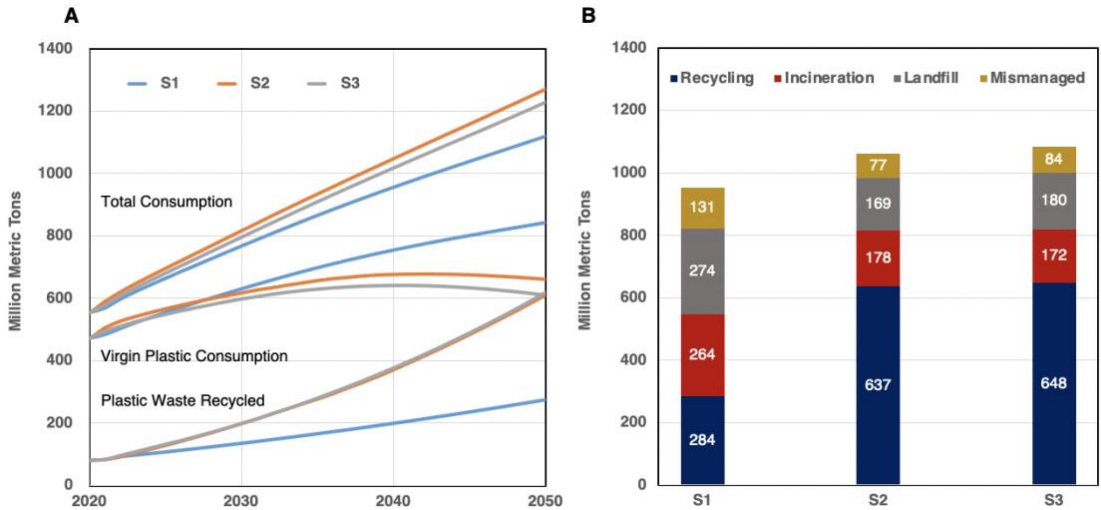

**Fig. 4. 2020-2050 scenarios overview. S1 – Business as Usual (BAU), S2 – 2050 60% recycling rate with 75% replacement of virgin plastic production, S3 – 2050 100% recycling rate for packaging only with 75% replacement of virgin packaging production.** (**A**) 2020-2050 influence of recycled waste on total and virgin plastic consumption. (**B**) 2050 global plastic end-of-life management under three scenarios.

12